# Grafted Low-Leakage Si/AlN p-n Diodes Enabled by Fluorinated AlN Interface

Yi Lu[1, #], Tsung-Han Tsai[1, #], Qingxiao Wang[2,†], Haicheng Cao[2, †], Jie Zhou[1], You Jin Koo[1], Chenyu Wang[1], Yang Liu[1], Yueyue Hao[1], Michael Eller[4], Connor Bailey[4], Stephanie Liu[4], Nicholas J. Tanen[4], Zhiyuan Liu[2], Mingtao Nong[2], Robert M. Jacobberger[1], Tien Khee Ng[2], Katherine Fountaine[4], Vincent Gambin[4,*], Boon S. Ooi[2,5], Xiaohang Li[2,*], and Zhenqiang Ma[1, *]

[1] Department of Electrical and Computer Engineering, University of Wisconsin-Madison, Madison, WI, 53706, USA

[2] Department of Electrical and Computer Engineering, King Abdullah University of Science and Technology, Thuwal 23955-6900, Saudi Arabia

[3] Department of Materials Science and Engineering, University of Wisconsin-Madison, Madison, WI, 53706, USA

[4] Northrop Grumman Corporation, Redondo Beach, CA 90278, USA

[5] Department of Electrical, Computer, and Systems Engineering, Rensselaer Polytechnic Institute, Troy, NY 12180, USA

[*]Authors to whom correspondence should be addressed: vincent.gambin@ngc.com xiaohang.li@kaust.edu.sa or mazq@engr.wisc.edu

[#] Those authors contributed equally to this work

[†]Those authors contributed equally to this work

**Abstract:**

Ultrawide-bandgap AlN is highly attractive for next-generation power electronic devices, yet its practical application is severely limited by unmanageable surface chemistry and the high activation energy of its p-type dopant. In particular, high-temperature rapid thermal annealing (RTA), which is essential for low-resistance contacts on n-type AlN, inevitably induces thick and defective surface oxides that degrade heterojunction performance. In this work, we demonstrate an effective interface engineering strategy based on fluorination-derived $AlF_x$ and SiNx passivation to suppress defect-assisted leakage in p-Si/n-AlN heterojunction PN diodes fabricated via a semiconductor grafting process. A low-damage pseudo-atomic layer etching (pseudo-ALE) process is first employed to remove the RTA-induced oxide and restore a near-stoichiometric AlN surface. Subsequent $XeF_2$ treatment introduces an ultrathin $AlF_x$ layer, which is then covered and stabilized by an atomic-layer-deposited $SiN_x$ capping layer prior to p-Si nanomembrane transfer. Electrical characterization reveals that the $AlF_x/SiN_x$-engineered interface reduces the reverse leakage current by several orders of magnitude compared with thermally oxidized and oxide-removed AlN surfaces, while maintaining comparable forward-bias conduction. Temperature-dependent measurements show that the Poole–Frenkel emission process dominating moderate-field leakage is strongly suppressed, and the onset of reverse-bias current increase is shifted to significantly higher reverse bias, which is solely limited by AlN crystal quality. X-ray photoelectron spectroscopy and transmission electron microscopy analyses confirm the formation of Al–F bonding, suppressed Al–O formation, and a thin $SiO_x$/SiON interfacial passivation layer. These results identify $AlF_x$/SiNx as an effective and robust passivation scheme for stabilizing AlN interfaces and provide a viable pathway toward low-leakage ultrawide-bandgap heterojunction devices.

Aluminum nitride (AlN), an ultrawide-bandgap (UWBG) semiconductor with a bandgap of ~6.1 eV and a theoretical breakdown field exceeding 12 MV cm$^{-1}$, holds great potential for next-generation high-power, high-frequency, and far-ultraviolet electronic devices[1–5]. In addition to its high thermal conductivity and chemical stability, AlN is also attractive as both an active material and a platform for heterogeneous integration with other semiconductors[6]. However, realizing high-performance AlN-based electronic devices remains challenging, largely due to the poor controllability of AlN surface and interface chemistry, which strongly affects carrier transport, interface traps, and leakage in devices such as metal–oxide–semiconductor field-effect transistors (MOSFETs), Schottky diodes, and heterojunctions[7–9].

The difficulty in forming high-quality interfaces on AlN originates from its intrinsically low electron affinity and high chemical reactivity[10–12], which make AlN surfaces prone to oxidation even under mild processing conditions. Previous studies have shown that annealing above ~1000 °C results in the formation of amorphous or partially crystallized AlOx layers on AlN surfaces[13–15]. While oxides such as $Al_2O_3$ are widely used for passivation in Si and GaN technologies, their application to AlN is problematic. The low conduction-band offset between $Al_2O_3$ and AlN, combined with oxygen-related defects, generates a high density of interface traps that promote trap-assisted tunneling and Fermi-level pinning[16,17]. For example, Liu et al. reported that surface oxides on Al-rich AlGaN contain a high density of defects that act as leakage paths in Schottky barrier diodes18.

This passivation challenge becomes more severe during device fabrication processes such as grafted heterojunction integration. Rapid thermal annealing (RTA), which is necessary for forming ohmic contacts on n-type AlN cathodes, inevitably induces strong oxidation at exposed AlN surfaces. The resulting thermally induced oxide is typically thick, partially crystalline, and electrically defective[18], containing numerous oxygen-related traps that form leakage paths and cause severe Fermi-level pinning in heterojunction devices. Our previous work revealed severe leakage at the Si/AlN interface governed by Poole–Frenkel emission at moderate bias and trap-assisted tunneling at high electric fields due to AlN surface oxidation[19]. Although buffered oxide etchant (BOE) cleaning can remove the amorphous oxide, it is ineffective against crystallized AlOx domains formed during high-temperature RTA[20]. Therefore, developing a low-damage strategy to remove defective oxides while stabilizing the AlN surface is essential for improving device reliability.

Fluorination provides a promising route to address this issue. Fluorine forms strong Al–F bonds that effectively passivate Al dangling bonds and suppress oxygen adsorption[21,22]. The resulting AlF$_x$ layer is expected to provide improved interfacial stability compared with conventional AlOx. In addition, AlF$_3$ has been reported to exhibit a larger conduction-band offset with AlN than $Al_2O_3$, suggesting that AlF$_3$ (or AlF$_x$) may serve as a more effective dielectric for reducing leakage and interface traps[23,24]. Moreover, Al–F bonds possess higher bond energy than Al–O bonds, providing enhanced resistance to reoxidation during device processing[25].

In this work, we systematically investigate the influence of controlled fluorination and surface chemistry modification on the electrical behavior of p-Si/n-AlN heterojunctions fabricated through a semiconductor grafting process. A low-power pseudo-atomic layer etching (pseudo-ALE) process is first employed to remove the partially crystallized thermal oxide and expose a stoichiometric AlN surface, followed by XeF$_2$ fluorination to form an AlF$_x$ passivation layer. By comparing thermally oxidized, chemically cleaned, and fluorinated AlN surfaces, we establish a direct correlation between surface composition, interface defect density, and diode leakage characteristics. These results highlight the importance of surface chemistry engineering for stabilizing AlN interfaces and demonstrate $AlF_x$/SiN as an effective passivation strategy for AlN-based ultrawide-bandgap semiconductor devices.

In this study, an ~800 nm n-type AlN layer was grown on (0001) sapphire substrates by metal–organic chemical vapor deposition (MOCVD) using trimethylaluminum (TMA) and ammonia ($NH_3$) precursors. The Si doping concentration was ~$2 \times 10^{18}$ cm$^{-3}$ with a carrier concentration of ~$10^{15}$ cm$^{-3}$. After cleaning, Ti/Al/Ti (20/120/80 nm) metal pads were patterned by photolithography and electron-beam evaporation, followed by rapid thermal annealing (RTA) at 1100 °C for 30 s in $N_2$ to form low-resistance cathodes.

To remove the RTA-induced oxide, a low-power ICP-RIE process using $BCl_3$/$Cl_2$ plasma was applied for 5 min, yielding an etch rate of ~1.8 nm min$^{-1}$. This process is referred to as pseudo-atomic layer etching (pseudo-ALE). The samples were subsequently immersed in 10:1 BOE for 10 min, rinsed with deionized water, and dried under $N_2$ before fluorination. Surface fluorination was carried out in a $XeF_2$ etcher at room temperature with 30 s pulses per cycle. Treatments of 10, 30, and 60 cycles were examined to optimize fluorination. Immediately after $XeF_2$ treatment, an ultrathin $SiN_x$ layer (~0.5 nm, 33 cycles) was deposited by atomic layer deposition using bis(diethylamino)silane (BDEAS) and $N_2$ plasma at 300 °C to stabilize the fluorinated surface.

A ~180 nm boron-doped (~$5 \times 10^{19}$ cm$^{-3}$) p-type Si nanomembrane (NM) was released from a p-type SOI wafer. The top Si layer was patterned with 9 μm × 9 μm grid openings to expose the buried oxide, which was selectively etched in 49% HF for ~4 h to release the Si NM. The membrane was then picked up using a PDMS stamp and transferred onto the prepared n-AlN surface via the semiconductor grafting process. After alignment, the stack was annealed at 350 °C for 5 min in $N_2$ to form chemical bonding at the Si/$AlF_x$/$SiN_x$/Al interface. Following transfer, Ni/Au/Cu/Au (10/10/100/10 nm) anodes were deposited on the p-Si surface by electron-beam evaporation. Mesa isolation was performed by RIE using the anode metal as a hard mask. Three device structures were fabricated for comparison: (A) RTA-oxidized AlN with thermal oxide, (B) pseudo-ALE + BOE-cleaned AlN surface, and (C) $XeF_2$-fluorinated AlN with an $AlF_x$/$SiN_x$ interfacial layer.

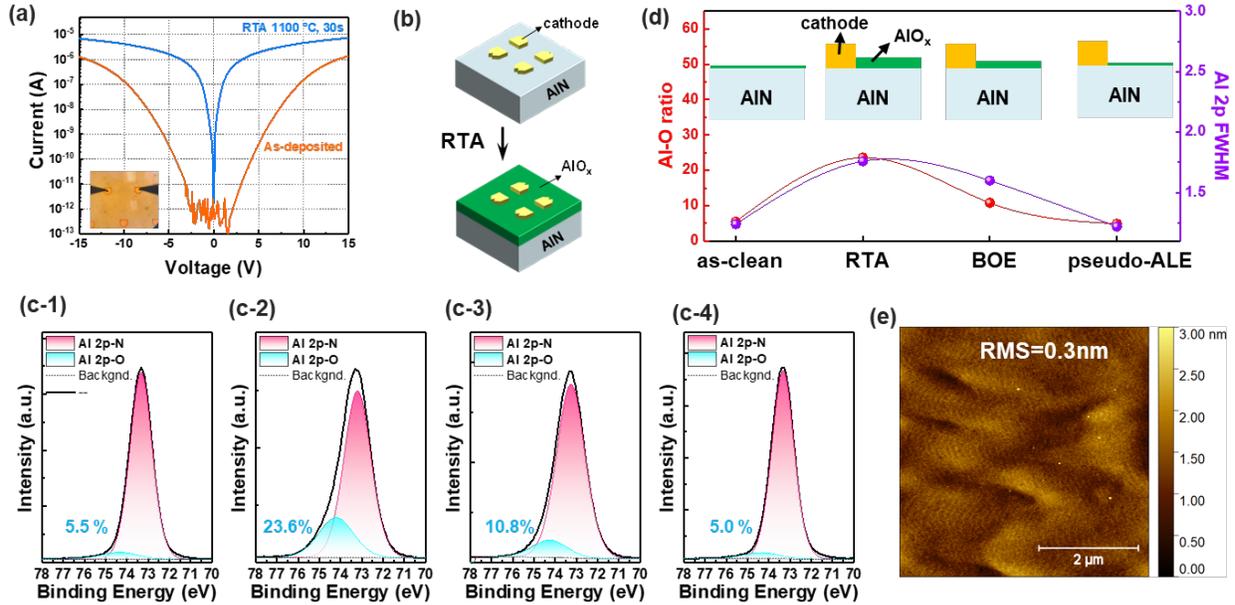

Figure 1. (a) IV curves between two metal pads on n-AlN with and without RTA, inset is the microscope image of the metal pad on n-AlN; (b) Illustration of AlN surface after cathode annealing. (c) Fitted Al 2p XPS spectra after after (c-1) BOE cleaning, (c-2) RTA 1100 °C 30s. (c-3) BOE cleaning again, (c-4) pseudo-ALE etching; (d) Al-O/ (Al-N+Al-O) ratio and Al 2p FWHM values change during step c-1 to c-4 process. (e) AlN surface morphology after (c-4) process.

To reduce the contact resistance, an optimized RTA condition (1100 °C for 30 s in $N_2$ ambient) was employed after cathode metallization[26]. Figure 1a shows the I–V characteristics measured between two Ti/Al/Ti metal pads on as-grown and RTA-treated AlN surfaces. The RTA-treated sample exhibits significantly higher current, indicating reduced contact resistance. However, even under $N_2$ ambient, high-temperature RTA can induce severe AlN surface oxidation due to the strong reactivity of aluminum with trace $O_2/H_2O$ impurities that react readily with the Al-terminated surface (Figure. 1b).

As shown in Figure 1c, the BOE-cleaned surface exhibits a dominant Al–N component with a small Al–O ratio of ~5.5% (Figure 1c-1). After RTA at 1100 °C for 30 s, the Al–O component increases significantly to 23.6% (Figure 1c-2), accompanied by broadening of the Al 2p FWHM (Figure 1d), indicating increased chemical disorder. Even after an additional 10 min BOE cleaning, a considerable Al–O component remains (10.8%), suggesting that the RTA-induced oxide partially contains crystallized $AlO_x$ that cannot be removed by BOE.

To remove this defective oxide, a low-power etching process (pseudo-ALE, see Experimental Section) was employed, removing ~9 nm of surface material and selectively eliminating the Al–O layer without damaging the underlying AlN (Figure 1c-4). The treated surface shows reduced Al 2p FWHM and recovery of the Al–O ratio to ~5%, indicating a cleaner and more stoichiometric Al–N surface. AFM measurement further shows a smooth morphology with a roughness of ~0.3 nm (Figure 1e), confirming that the mild plasma process does not introduce additional surface damage. These results demonstrate that pseudo-ALE effectively restores a uniform AlN surface suitable for subsequent interface engineering.

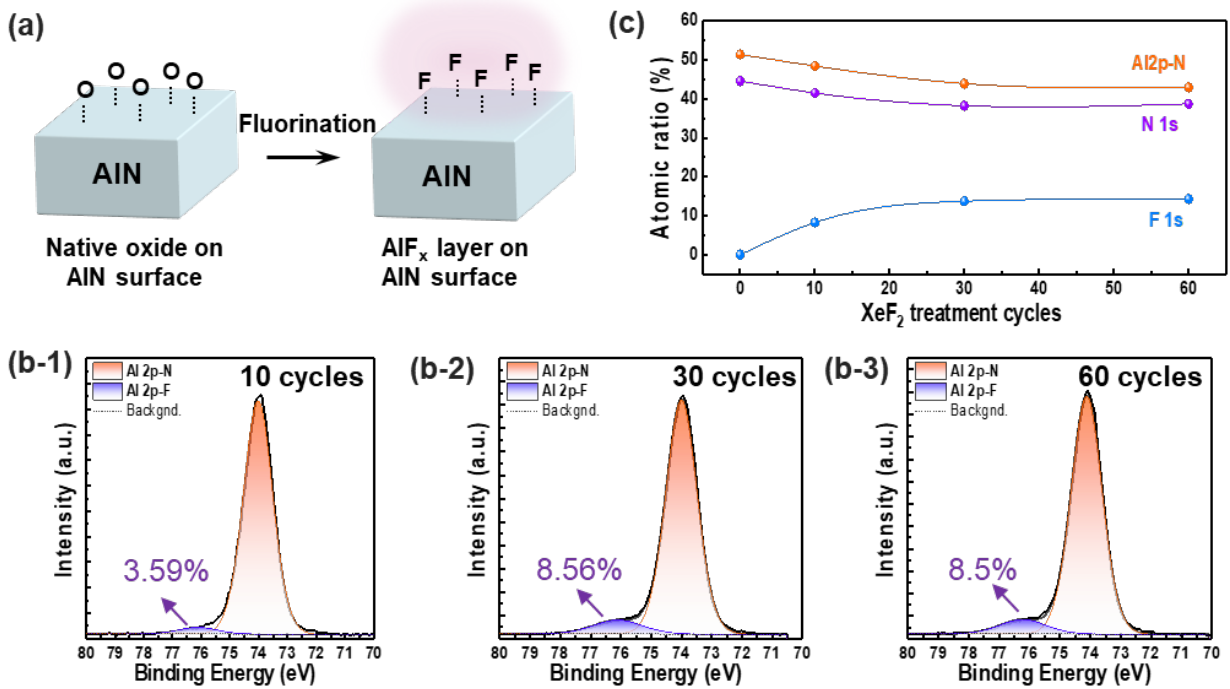

Figure 2. (a) Illustration of fluorination process on AlN surface; Al 2p spectra of AlN surface after XeF2 treatment of (b-1) 10 cycles, (b-2) 30 cycles, and (b-3) 60 cycles; (c) extracted atomic ratio of Al 2p-N, N 1s, F 1s under different $XeF_2$ treatment conditions.

After removal of the thermal oxide, $XeF_2$ treatment was applied to modify the chemical state of the AlN surface. As illustrated in Figure 2a, this process replaces surface Al–O bonds with Al–F bonds, forming an ultrathin $AlF_x$ passivation layer. The Al 2p spectra in Figures 2b-1 to 2b-3 show progressive formation of Al–F bonds with increasing $XeF_2$ cycles. As the cycle number increases from 10 to 60, the Al–F shoulder becomes more pronounced. Quantitative analysis indicates that the $AlF_x/(AlF_x + AlN)$ fraction increases from 3.59% (10 cycles) to 8.56% (30 cycles) and then saturates at ~8.5% (60 cycles). Consistently, the F 1s atomic ratio increases rapidly up to 30 cycles and then saturates (Figure 2c), indicating completion of the fluorination reaction.

The formation of Al–F bonds significantly improves interface stability. Fluorination passivates Al dangling bonds and suppresses oxygen adsorption, since the Al–F bond is stronger than Al–O (≈660–680 kJ/mol vs. ≈500–520 kJ/mol)[27]. In addition, the $AlF_x$ layer forms a chemically inert surface that resists reoxidation. However, the $AlF_x$ surface alone shows limited mechanical robustness and adhesion for subsequent p-Si nanomembrane transfer. To stabilize the interface, an ultrathin $SiN_x$ layer (~0.5 nm, 33 cycles) was deposited by ALD, forming an $AlF_x/SiN_x$ interfacial layer. The $SiN_x$ layer (i) seals the fluorinated AlN surface from ambient oxygen and moisture, (ii) passivates surface states on both the AlN and Si sides, and (iii) mechanically stabilizes the interface for successful p-Si nanomembrane transfer. The resulting $AlF_x/SiN_x$ bilayer therefore provides both electronic and mechanical stabilization for forming reproducible low-leakage AlN heterojunctions.

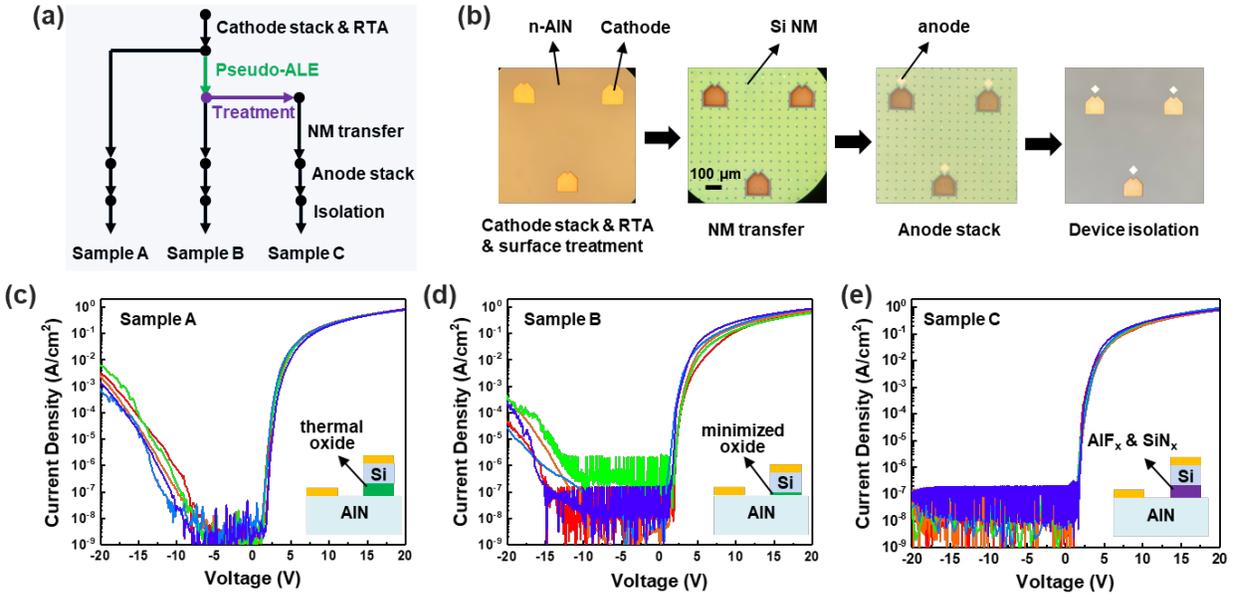

Figure 3. (a) Fabrication flow of Sample A, B, C; (b) Microscope images during the fabrication process for Sample C; typical IV curves of (c) Sample A (d) Sample B and (e) Sample C (five devices for each).

Using the three types of AlN surfaces described above: Sample A: thermal-oxide interface; Sample B: pseudo-ALE-treated interface with minimized surface oxide; Sample C: $AlF_x/SiN_x$ interface; p-Si/n-AlN heterojunction diodes were fabricated and compared (Figure 3a). The fabrication process is shown in Experimental section, and the fabrication images for sample C are shown in Figure 3b. Representative I–V curves of five devices for each sample are shown in Figures 3c–3e. All three types of diodes exhibit clear rectifying characteristics, confirming the formation of a p–n heterojunction. However, distinct differences are observed in reverse leakage current. In sample A (thermal-oxide interface) severe leakage current occurs at -5 V - -7.5V, which is attributed to abundant oxygen-related defect states and local crystalline $AlO_x$ grains within the thick oxide layer formed during high-temperature RTA. These defects provide multiple leakage paths, leading to large leakage current among devices.

Sample B (pseudo-ALE-treated interface) also exhibits pronounced reverse leakage (around -10 to -15 V) and non-uniform device behavior. The large leakage is mainly attributed to unsaturated dangling bonds and native vacancy defects remaining on the AlN surface after oxide removal[28,29]. Although pseudo-ALE effectively eliminates the thermally induced oxide, it exposes a chemically reactive Al-terminated surface containing abundant dangling bonds and vacancy-related defects[30]. As a result, the unpassivated AlN surface remains electronically unstable, leading to significant leakage and device variability. In addition, AlN surfaces rapidly reoxidize after etching because the Al-terminated surface readily reacts with residual $O_2$ and $H_2O$[13,14]. This reoxidation is often spatially non-uniform due to variations in surface polarity, defect density, and local stoichiometry, producing inhomogeneous interface states and localized leakage paths, which account for the device-to-device variability observed in Figure 3d.

In contrast, Sample C ($AlF_x/SiN_x$ interface) exhibits the lowest reverse current and excellent device uniformity up to -20 V. The I–V curves in Figure 3e show a significant leakage reduction compared with Samples A and B. Statistical analysis of reverse/forward current (Figure 4a), on/off ratio (Figure 4b), and $R_{ON}$ (Figure 4c) further confirms the improved reproducibility and performance of Sample C. The reverse current density at -20 V is ~$10^{-8}$ A cm$^{-2}$, which is 4-6 orders of magnitude lower than that of Samples A and B. At ±20 V, Sample C shows an average rectification ratio of ~$3 \times 10^7$, whereas Samples A and B exhibit values of ~475 and ~$10^4$, respectively. Similar $R_{ON}$ values indicate comparable forward transport, suggesting that the reverse leakage differences mainly originate from interface-state-mediated transport.

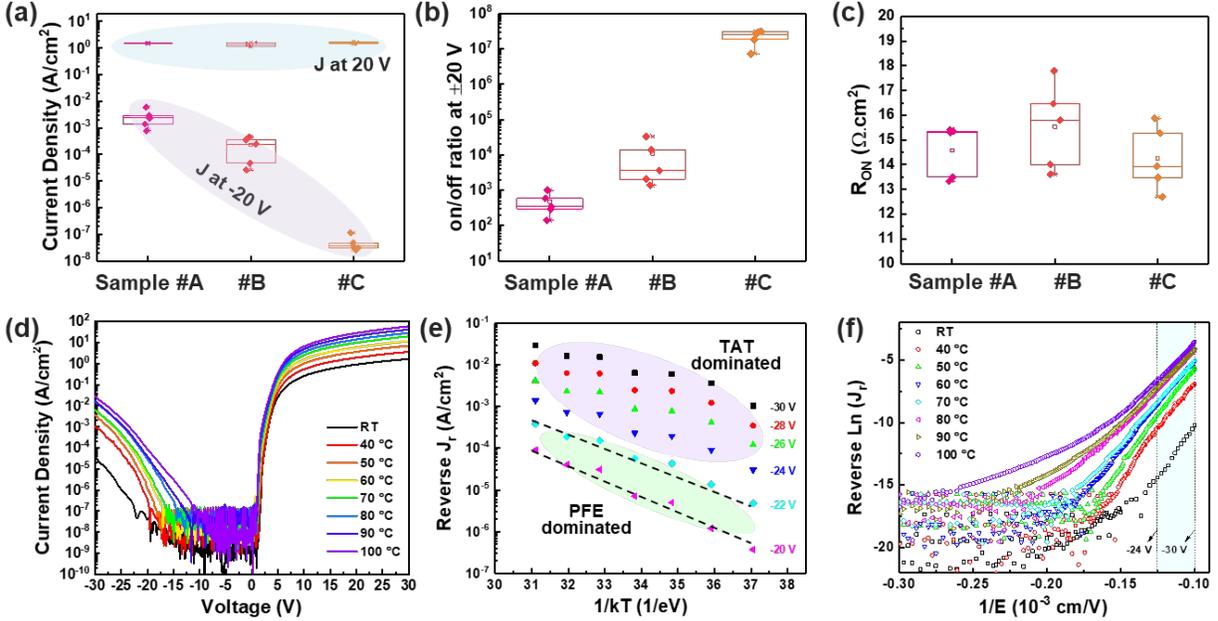

Figure 4. (a) Extracted reverse-bias current at -20 V, (b) on/off current ratio (at ±20 V), and (c) specific on-resistance ($R_{ON}$) of Samples A–C. (d) Temperature-dependent I–V characteristics of Sample C measured from room temperature (RT) to 100 °C. (e) Arrhenius plot of reverse current density ($J_r$) as a function of 1/kT. (f) ln($J_r$) plotted as a function of 1/E, illustrating the field-dependent transport behavior at high reverse biases.

To clarify the origin of the improved performance of Sample C, temperature-dependent I–V measurements were performed from RT to 100 °C with a bias range of -30 V to 30 V. The I–V curve of a representative device is shown in Figure 4d. As the temperature increases from RT to 100 °C, both forward and reverse currents increase, indicating a thermally activated transport process. In our previous study of p-Si/n-AlN PN diodes with inadequately treated interfaces (Sample A), the reverse leakage current was governed by Poole–Frenkel emission (PFE)[31,32] at moderate reverse biases (-5 to -18 V) and trap-assisted tunneling (TAT) at higher electric fields (-18 to -30 V)[19].

For Sample C, the reverse current still exhibits temperature-dependent onset behavior (Figure 4d), but the onset of noticeable leakage shifts to significantly higher reverse voltages compared with Sample A, indicating suppression of defect-assisted transport at moderate electric fields. Further analysis in Figure 4e shows that the ln(Jr) versus 1/kT plots exhibit partial linearity at -22 V and -24 V, suggesting that thermally activated processes remain but are substantially weakened. This contrasts with Sample A, where a pronounced PFE-dominated regime was observed over -5 to -18 V. The reduced temperature sensitivity indicates that the density and/or activity of charged trap states responsible for the PFE process is significantly reduced by the interface engineering.

At higher reverse biases (|V| ≥ ~24 V), the leakage current becomes increasingly field-dominated, as evidenced by the good linearity in the ln(Jr) versus 1/E plot in Figure 4f together with weak temperature dependence in Figure 4e. This behavior is characteristic of a TAT process[33,34]. Importantly, the transition to TAT occurs at significantly higher reverse voltages in Sample C, indicating delayed activation due to the reduced interface trap density and suppressed local electric-field enhancement. The remaining TAT is therefore attributed mainly to the dislocation density of the epitaxial AlN rather than interface defects. These results demonstrate that the reduced reverse leakage current in Sample C originates primarily from suppression of the PFE-dominated transport regime, while the TAT process is shifted to higher reverse bias due to improved interface quality.

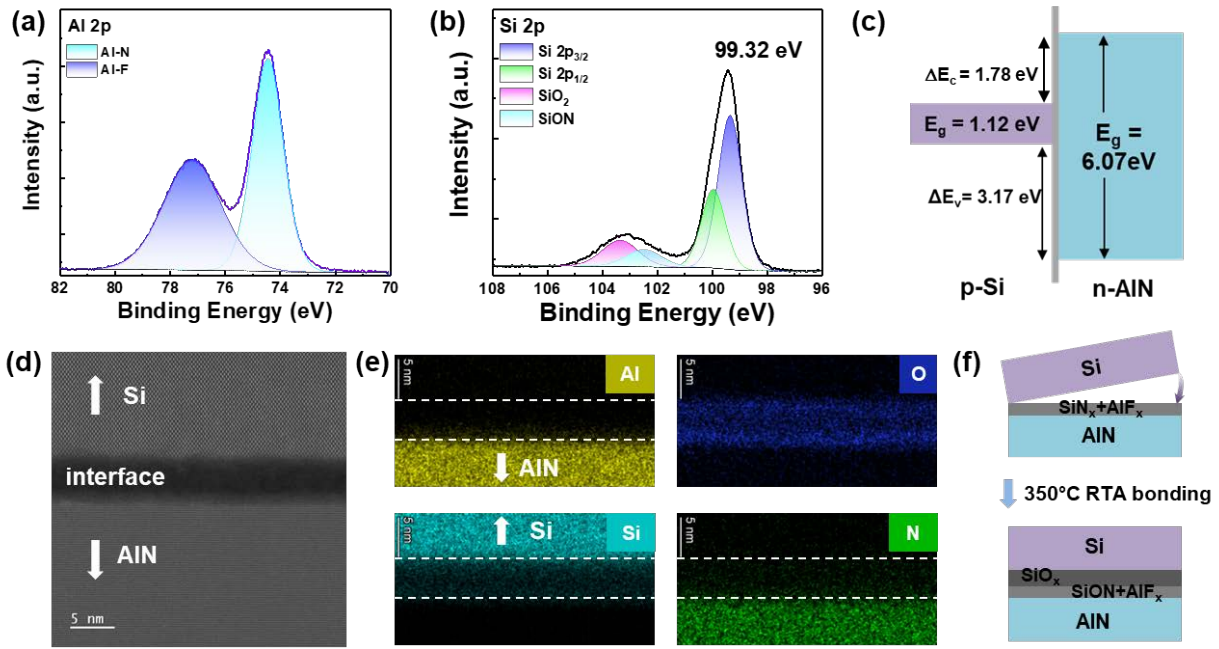

Figure 5. (a) XPS Al 2p and (b) Si 2p core-level spectra acquired at the Si/AlN interface of Sample C. (c) Reconstructed band alignment of the Si/AlN heterojunction based on the XPS analysis. (d) Cross-sectional transmission electron microscopy (TEM) image of the Si/AlN interface. (e) Corresponding energy-dispersive X-ray spectroscopy (EDS) elemental maps of the Si/AlN interface. (f) Schematic illustration of the Si/AlN interfacial structure before and after RTA bonding. Note: The F element amount at interface is too low to be imaged via EDS (*i.e.*, indistinguishable from background noise).

Figure 5 presents the interfacial chemical states, microstructure, and reconstructed band alignment of the Si/AlN interface in Sample C. As shown in Figure 5a, the Al 2p spectrum exhibits an additional high–binding-energy component besides the Al–N peak, attributed to Al–F bonding induced by XeF$_2$ fluorination. During the subsequent ALD deposition of an ultrathin SiN$_x$ layer at ~300 °C, these Al–F bonds are further stabilized. The Si 2p spectrum in Figure 5b shows the coexistence of SiO$_x$ and Si–O–N (SiON) components. The SiO$_x$ originates from native oxidation of the transferred p-Si nanomembrane prior to bonding, whereas the SiON phase results from oxygen diffusion into SiN$_x$ during the post-bonding RTA at 350 °C for 5 min. The reconstructed band alignment in Figure 5c therefore reflects the combined effects of the SiO$_x$/SiON interlayer and the Al–F–modified AlN surface.

Cross-sectional TEM imaging (Figure 5d) confirms that both the Si nanomembrane and AlN remain single crystalline, with a chemically intermixed interfacial region of ~5 nm formed during bonding. Elemental EDS mapping (Figure 5e) shows that the AlN side is largely free of oxidation, consistent with fluorination-induced suppression of Al–O formation. Two interfacial layers are observed near the Si side, corresponding to a $SiO_x$-rich region and an underlying SiON layer. This interface chemistry differs significantly from our previous Si/AlN heterojunction with native oxide and $Al_2O_3$ layers rich in Al–O bonds.[19] Although fluorine is detected by XPS, it cannot be resolved by EDS due to its low concentration. The interfacial evolution before and after bonding is illustrated in Figure 5f, showing the transition from the initial $SiN_x/AlF_x$-modified AlN interface to the final $SiO_x/SiON/AlF_x$ heterointerface that suppresses defect-assisted leakage in Sample C.

In summary, we demonstrate that fluorination-derived $AlF_x$ interface passivation effectively suppresses defect-assisted leakage in p-Si/n-AlN heterojunction PN diodes. Although high-temperature RTA is required to form low-resistance n-AlN contacts, it inevitably generates a thick and electrically defective surface oxide that causes severe reverse leakage through Poole–Frenkel emission and trap-assisted tunneling. By combining a low-damage pseudo-ALE process to remove the RTA-induced oxide with $XeF_2$ fluorination, a chemically stable $AlF_x$-modified AlN surface is achieved. An ultrathin ALD $SiN_x$ layer further stabilizes the interface and enables effective passivation of the transferred p-Si nanomembrane. Compared with thermally oxidized and oxide-removed interfaces, the $AlF_x/SiN_x$-engineered interface exhibits several orders-of-magnitude reduction in reverse leakage current and improved device uniformity while maintaining similar forward conduction. Temperature-dependent analysis indicates that the dominant leakage mechanism shifts from Poole–Frenkel emission to a weaker thermally activated process, with the onset of trap-assisted tunneling delayed to higher reverse bias. These results establish a direct correlation between AlN surface chemistry, interface defect states, and leakage transport, highlighting $AlF_x$ passivation as a promising strategy for AlN-based ultrawide-bandgap devices.

**Acknowledgments**

This work was supported by DARPA UWBGS program under award 140D04-24-C-0061 and by National Science Foundation with award number ECCS 2235377, KAUST Baseline Fund BAS/1/1664-01-01. Y.J.K. and R.M.J. were supported by the U.S. Department of Energy, Office of Science, Basic Energy Sciences, Grant DE-SC0026046. The views, opinions and/or findings expressed are those of the authors and should not be interpreted as representing the official views or policies of the Department of War or the U.S. Government. DARPA Distribution Statement "A" (Approved for Public Release, Distribution Unlimited).

**Conflict of interest**

The authors declare no conflict of interest.

**Data availability statement**

The data that support the findings of this study are available from the corresponding author upon reasonable request.